\documentclass[twocolumn,showpacs,preprintnumbers,amsmath,amssymb]{revtex4}

\usepackage[dvips]{graphicx}

\begin{document}
\title{Leptogenesis with supersymmetric Higgs triplets in the TeV region}

\author{Masato Senami${}^{\rm a,b}$}
\altaffiliation{E-mail address: senami@icrr.u-tokyo.ac.jp; \\
b: address after April, 2004}
\author{Katsuji Yamamoto${}^{\rm a}$}
\altaffiliation{E-mail address: yamamoto@nucleng.kyoto-u.ac.jp}
\affiliation{
${}^{\rm a}$Department of Nuclear Engineering, Kyoto University,
Kyoto 606-8501, Japan \\
${}^{\rm b}$Institute for Cosmic Ray Research, University of Tokyo,
Kashiwa City, Chiba 277-8582, Japan}

\date{\today}

\begin{abstract}
The leptogenesis with supersymmetric Higgs triplets is studied
in the light of experimental verification in the TeV region.
The lepton number asymmetry appears just after the inflation
via multiscalar coherent evolution of Higgs triplets and antislepton
on a flat manifold.
If the Higgs triplet mass terms dominate over the negative thermal-log term
for the Hubble parameter $ H $
comparable to the Higgs triplet mass $ M_\Delta $,
the asymmetry is fixed readily to some significant value
by the redshift and rotation of these scalar fields,
providing the sufficient lepton-to-entroy ratio $ n_L / s \sim 10^{-10} $.
This can be the case even with $ M_\Delta \sim 1 {\rm TeV} $
for the reheating temperature $ T_{\rm R} \sim 10^6 {\rm GeV} $
and the mass parameter $ M / \lambda \sim 10^{22} {\rm GeV} $
of the nonrenormalizable superpotential terms relevant for leptogenesis.
\end{abstract}

\pacs{14.80.Cp, 98.80.Cq, 14.60.Pq}
\keywords{leptogenesis, neutrino mass, Higgs triplet, supersymmetry}
\maketitle

%%%%%%%%%%%%%%%%%%%%
%%%%%%%%%%%%%%%%%%%%
\section{Introduction}
\label{sec:introduction}
%%%%%%%%%%%%%%%%%%%%
%%%%%%%%%%%%%%%%%%%%

The baryogenesis is a very important subject
in particle physics and cosmology.
In most of the baryogenesis scenarios including those of leptogenesis,
however, the participating particles are supposed to be extremely heavy,
and hence it seems impossible to verify them experimentally.
The electroweak baryogenesis might be promising in this respect,
though it is realized in a rather restricted situation
within the minimal supersymmetric standard model.

In this paper we study the leptogenesis with supersymmetric Higgs triplets
in the light of experimental verification in the TeV region.
It is indeed interesting that the neutrino masses may be generated
naturally by the exchange of Higgs triplet
\cite{higgstriplet}.
In particular, phenomenological implications
of the Higgs triplets with mass
$ M_\Delta \sim 100 {\rm GeV} - 1 {\rm TeV} $,
such as lepton flavor violating processes,
have been investigated recently, which are intimately related
to the neutrino masses and mixings
\cite{t3ha,t3hb,t3hc,PRD69,rossi}.
It is hence attractive to study the possibility of leptogenesis
with Higgs triplets in the TeV region.
In this respect we note that in the supersymmetric Higgs triplet model
the leptogenesis can be realized after the inflation
via multiscalar coherent evolution on a flat manifold
of a pair of Higgs triplets $ \Delta $, $ {\bar \Delta} $
and the antislepton $ {\tilde e}^c $
\cite{PLB524,PRD66-67}
in the manner of Affleck and Dine
\cite{AD,DRT}.
The Higgs triplet mass was originally
supposed to be very large as $ M_\Delta \sim 10^{9}-10^{14} {\rm GeV} $
\cite{PLB524},
so that it provides a strong driving force
for the rotation of scalar fields to fix the lepton number asymmetry
to some significant value.
We here reexamine this scenario to show that the successful leptogenesis
can be obtained even with $ M_\Delta \sim 1 {\rm TeV} $,
where the asymmetry fixing becomes a more important issue
due to the effects of thermal terms
\cite{thermaleffect1,thermaleffect2}.

This paper is organized as follows.  In Sec. \ref{sec:model},
we present the supersymmetric Higgs triplet model,
and describe the neutrino mass generation.
In Sec. \ref{sec:asymmetry}, we recapitulate the essential aspects
of the generation of lepton number asymmetry after the inflation
via coherent evolution of $ \Delta $, $ {\bar \Delta} $, $ {\tilde e}^c $
on the flat manifold.
In Sec. \ref{sec:completion}, we examine the completion of leptogenesis
to show that the sufficient lepton-to-entroy ratio
$ n_L / s \sim 10^{-10} $ can be obtained
even for the case of phenomenologically interesting Higgs triplets
with $ M_\Delta \sim 1 {\rm TeV} $.
The thermal terms provide significant effects
on the evolution of $ \Delta $, $ {\bar \Delta} $, $ {\tilde e}^c $
and the lepton number asymmetry
for the Hubble parameter $ H < M_\Delta $.
Section \ref{sec:summary} is devoted to the summary.

%%%%%%%%%%%%%%%%%%%%
%%%%%%%%%%%%%%%%%%%%
\section{Model}
\label{sec:model}
%%%%%%%%%%%%%%%%%%%%
%%%%%%%%%%%%%%%%%%%%

We investigate an extension of the minimal supersymmetric standard model
by introducing a pair of Higgs triplet superfields,
%%%%%
\begin{eqnarray}
\Delta & = & \left( \begin{array}{cc}
\Delta^+ / {\sqrt 2} & \Delta^{++} \\
\Delta^0 & - \Delta^+ / {\sqrt 2} \end{array} \right) ,
\\
{\bar \Delta} & = & \left( \begin{array}{cc}
{\bar \Delta}^- / {\sqrt 2} & {\bar \Delta}^0 \\
{\bar \Delta}^{--} & - {\bar \Delta}^- / {\sqrt 2} \end{array} \right) .
\end{eqnarray}
%%%%%
The lepton doublets $ L_i = ( \nu_i , l_i ) $,
anti-lepton singlets $ l_i^c $ ($ i = 1, 2, 3 $)
and the Higgs doublets $ H_u $, $ H_d $ are given as usual.
The generic lepton-number conserving superpotential
for the leptons and Higgs fields is given by
%%%%%
\begin{eqnarray}
W_0 = h_{ij} L_i H_d l^c_j + \mu H_u H_d 
    + \frac{1}{\sqrt 2} f_{ij} L_i \Delta L_j
    + M_\Delta {\bar \Delta} \Delta .
\label{W0}
\end{eqnarray}
%%%%%
The lepton numbers are assigned to the Higgs triplets as
%%%%%
\begin{eqnarray}
Q_L ( \Delta ) = -2 , \ Q_L ( {\bar \Delta} ) = 2 .
\end{eqnarray}
%%%%%
The lepton-number violating terms may also be included
in the superpotential as
%%%%%
\begin{eqnarray}
W_{\rm LV} = \xi_1 {\bar \Delta} H_u H_u + \xi_2 \Delta H_d H_d .
\label{WLV}
\end{eqnarray}
%%%%%%%%%%%%%%%%%%%%
These terms are $ R $-parity conserving
by assigning the Higgs triplets to be $ R $-parity even.
We do not consider $ R $-parity violating terms for definiteness.

The Higgs triplets develop nonzero vacuum expectation values
(VEV's) due to the effects of $ W_{\rm LV} $ as
%%%%%
\begin{eqnarray}
\langle {\Delta}^0 \rangle
= - c_1 \frac{\xi_1 \langle H_u \rangle^2 }{M_\Delta} ,
\langle {\bar \Delta}^0 \rangle
= - c_2 \frac{\xi_2 \langle H_d \rangle^2 }{M_\Delta} .
\label{tripletvev}
\end{eqnarray}
%%%%%
The factors $ c_1 , c_2 \sim 1 $ with $ \xi_1 \sim \xi_2 $
are determined precisely by minimizing the scalar potential
including the soft supersymmetry breaking terms
with the mass scale $ m_0 \sim 10^3 {\rm GeV} $
($ c_1 = c_2 = 1 $ in the supersymmetric limit of $ m_0 \rightarrow 0 $
with the vanishing $ F $ terms).
The neutrino mass matrix is then generated by the VEV of the Higgs triplet
\cite{higgstriplet} as
%%%%%
\begin{equation}
M_{\nu} = f {\sqrt 2} \langle \Delta^0 \rangle .
\end{equation}
%%%%%
This neutrino mass matrix should reproduce
the masses $ m_i $ and mixing angles $ \theta_{ij} $
inferred from the data of neutrino experiments
\cite{SK,SNO,CHOOZ,KamLand}.
The constraint on the magnitude of $ f $ coupling
is then placed from $ m_i \lesssim 10^{-1} {\rm eV} $ roughly as
%%%%%
\begin{eqnarray}
| f | \lesssim 10^{-1}
\left( \frac{\xi}{10^{-10}} \right)^{-1}
\left( \frac{M_\Delta}{10^3 {\rm GeV}} \right) .
\label{fcond}
\end{eqnarray}
%%%%%
This constraint, however, does not seem so stringent,
allowing even $ | f | \sim 1 $ and $ M_\Delta \sim 1 {\rm TeV} $
with small enough $ \langle \Delta^0 \rangle $.
Then, the interesting phenomenology of lepton flavor violation
are provided intimately related to the neutrino masses and mixings
\cite{t3ha,t3hb,t3hc,PRD69,rossi}.

The generation of the very small VEV's of the Higgs triplets
in Eq. (\ref{tripletvev}) for the neutrino masses
has been described in the literature
in terms of the trilinear couplings of Higgs doublets and triplets,
as given in Eq. (\ref{WLV}),
to break explicitly the lepton-number conservation
\cite{higgstriplet,t3hb,PLB524,PRD69,extra-dimension}.
Specifically, for the Higgs triplets in the TeV region
with $ M_\Delta \sim 1 {\rm TeV} $, the small VEV's of the Higgs triplets
are stably generated by the tiny couplings
$ \xi_1 \sim \xi_2 \sim \xi \sim 10^{-10} $,
as given in Eq. (\ref{fcond}).
We recapitulate below the essential points of this feature,
while it is not directly related to the present scenario of leptogenesis,
which is described in the following sections.

In the absence of the trilinear couplings $ \xi_1 $ and $ \xi_2 $,
the stable lepton-number conserving minimum
with the vanishing VEV's $ \Delta = {\bar \Delta} = 0 $
are generated as usual in a reasonable range of model parameters,
since the Higgs triplet mass terms
$ M_\Delta^2 ( | \Delta |^2 + | {\bar \Delta} |^2 ) $
with $ M_\Delta \sim 1 {\rm TeV} $
may dominate over the soft supersymmetry breaking terms with $ m_0 $.
This feature is still valid even if the radiative corrections
are included in the effective scalar potential.
Then, by adding the small lepton-number violating terms of $ W_{\rm LV} $,
the effective linear terms,
$ \xi_2 M_\Delta {\bar \Delta}^* \langle H_d \rangle^2 $,
$ \xi_1 m_0 {\bar \Delta}^0 \langle H_u \rangle^2 $, etc.,
are provided for the Higgs triplets.
The lepton-number violating part of the radiative corrections
are also small, since they should be generated
with the original $ \xi_1 $ and $ \xi_2 $ couplings.
Accordingly the potential minimum is shifted very slightly
by these terms to provide the small VEV's of the Higgs triplets
in Eq. (\ref{tripletvev}) breaking the lepton-number conservation.
It should be mentioned here that the VEV's of the Higgs triplets
are induced by the explicit lepton-number violation
of the $ \xi_1 $ and $ \xi_2 $ couplings,
rather than the spontaneous violation.
This is analogous to the case of explicit $ R $-parity violation
with the $ L H_u $ term, where the small VEV's of sleptons are induced.
The so-called triplet Majoron does not appear in the present case,
and all the scalar fields in the Higgs triplets
acquire masses $ \simeq M_\Delta $.
It is also seen that the slepton fields $ {\tilde L}_i $, $ {\tilde l}^c_i $
do not develop VEV's since the $ R $-parity is not violated
by the VEV's of the Higgs triplets.

In this way, the small VEV's of the Higgs triplets are attributed
to the small lepton-number violating trilinear couplings,
while keeping the Higgs triplet mass terms
($ M_\Delta \sim 1 {\rm TeV} $) dominant in the scalar potential.
This is clearly in contrast with the Coleman-Weinberg type potential.
In the Coleman-Weinberg case, while a very small VEV of a scalar field
is obtained at the tree-level
by tuning the mass term to be almost vanishing,
it is upset by the radiative corrections.
In the present model, although the lepton-number violating couplings
should be hierarchically small
to obtain the tiny VEV's of the Higgs triplets,
it is at least technically natural in the sense of 't Hooft \cite{natural}
against the radiative corrections 
with the dominant Higgs triplet mass terms at the tree-level.
This does not invoke any fine tining among the relevant couplings.
In the limit of $ \xi_1 , \xi_2 \rightarrow 0 $,
the global U(1) symmetry really appears for the lepton-number conservation,
and the VEV's of the Higgs triplets vanish.
The required tiny couplings $ \xi_1 \sim \xi_2 \sim \xi \sim 10^{-10} $
with $ M_\Delta \sim 1 {\rm TeV} $ and $ | f | \sim 1 $
may be understood more fundamentally by supposing
that the lepton-number violation originates in the Planck scale physics
\cite{PLB524,PRD69}.
It is also notable that the smallness of the Higgs triplet VEV's
may be explained in the context of large extra dimensions
\cite{extra-dimension}.

%%%%%%%%%%%%%%%%%%%%
%%%%%%%%%%%%%%%%%%%%
\section{Generation of asymmetry}
\label{sec:asymmetry}
%%%%%%%%%%%%%%%%%%%%
%%%%%%%%%%%%%%%%%%%%

We begin with recapitulating the essential aspects
of the generation of lepton number asymmetry
with supersymmetric Higgs triplets
via multiscalar coherent evolution on a flat manifold after the inflation
\cite{PLB524,PRD66-67}.
In the following, we consider for definiteness the case
that one generation of $ {\tilde e}^c $ ($ \equiv {\tilde l}^c_1 $),
together with $ \Delta $ and $ {\bar \Delta} $,
participates in the leptogenesis.
The essential results are even valid
for the case with more than one $ {\tilde l}^c $.
The nonrenormalizable terms relevant for leptogenesis are given by
%%%%%
\begin{eqnarray}
W_{\rm LG} = \frac{\lambda_{L \llap /}}{2M}
{\bar \Delta} {\bar \Delta} e^c e^c
+ {\rm e}^{i \delta}
\frac{\lambda_\Delta}{2M} {\bar \Delta} \Delta {\bar \Delta} \Delta ,
\label{Wnon}
\end{eqnarray}
%%%%%
where $ M $ represents some unification scale such as the Planck scale.
These terms represent the flat directions,
$ {\bar \Delta} {\bar \Delta} {\tilde e}^c {\tilde e}^c $ ($ Q_L = 2 $)
and $ {\bar \Delta} \Delta $ ($ Q_L = 0 $), respectively.
Then, if these directions are comparably flat with
%%%%%
\begin{equation}
0.3 \lesssim \lambda_{L \llap /} / \lambda_\Delta \lesssim 3 ,
\end{equation}
%%%%%
the coherent evolution of the scalar fields,
say AD-flatons \cite{AD,DRT,flaton},
may take place on the complex two-dimensional flat manifold
spanned by these directions, starting with large initial values
after the inflation.
This manifold is specified by the $ D $-flat condition,
%%%%%
\begin{eqnarray}
| \Delta^+ |^2 - | {\bar \Delta}^- |^2 + | {\tilde e}^c |^2 = 0 ,
\label{flat}
\end{eqnarray}
%%%%%
and the other fields are negligibly small.

The scalar potential for the AD-flatons is given by
%%%%%
\begin{eqnarray}
V
&=& ( C_1 m_0^2 - c_1 H^2 ) |\Delta|^2
 + ( C_2 m_0^2 - c_2 H^2 ) |{\bar \Delta}|^2
\nonumber \\
&& + ( C_3 m_0^2 - c_3 H^2 ) |{\tilde e}^c|^2
\nonumber \\
&& + \left| M_\Delta \Delta
 + \frac{\lambda_{L \llap /}}{M} {\bar \Delta} {\tilde e}^c {\tilde e}^c
 + {\rm e}^{i \delta}
   \frac{\lambda_\Delta}{M} {\bar \Delta} \Delta \Delta \right|^2
\nonumber \\
&& + \left| M_\Delta {\bar \Delta}
 + {\rm e}^{i \delta}
   \frac{\lambda_\Delta}{M} {\bar \Delta} {\bar \Delta} \Delta \right|^2
 + \left| \frac{\lambda_{L \llap /}}{M}
         {\bar \Delta} {\bar \Delta} {\tilde e}^c \right|^2
\nonumber \\
&& + \left[ ( b_\Delta H +  B_\Delta m_0 )
    M_\Delta {\bar \Delta} \Delta
 + {\rm h.c.} \right]
\nonumber \\
&& + \left[ \frac{1}{2M}
   ( a_{L \llap /} H + A_{L \llap /} m_0 )
    \lambda_{L \llap /}
    {\bar \Delta} {\bar \Delta} {\tilde e}^c {\tilde e}^c
 + {\rm h.c.} \right]
\nonumber \\
&& + \left[ \frac{1}{2M}
   ( a_\Delta H + A_\Delta m_0 )
   \lambda_\Delta {\bar \Delta} \Delta {\bar \Delta} \Delta
  + {\rm h.c.} \right]
\nonumber \\
&& + g_1^2 (|\Delta|^2 - |{\bar \Delta}|^2 + |{\tilde e}^c|^2)^2 ,
\label{V}
\end{eqnarray}
%%%%%
where the last term with the $ {\rm U(1)}_Y $ gauge coupling $ g_1 $
is included to realize dynamically the $ D $-flat condition (\ref{flat}).
(Henceforth $ \Delta \equiv \Delta^+ $
and $ {\bar \Delta} \equiv {\bar \Delta}^- $ for simplicity.)
The energy density of the universe
provides the soft supersymmetry breaking terms
with the Hubble parameter $ H $ \cite{DRT}.
The AD-flatons $ \phi_a = \Delta , {\bar \Delta} , {\tilde e}^c $
evolve in time governed by this potential $ V $.
Their number asymmetries are given with the time derivatives by
%%%%%
\begin{eqnarray}
n_a = i \left( \phi_a^* \dot{\phi}_a - \dot{\phi}_a^* \phi_a \right)
= \dot{\theta}_a | \phi_a |^2 ,
\end{eqnarray}
%%%%%
where
%%%%%
\begin{equation}
\phi_a (t) \equiv {\rm e}^{i \theta_a (t)} | \phi_a (t) | .
\end{equation}
%%%%%
Then, the lepton number asymmetry is evaluated by
%%%%%
\begin{equation}
n_L = 2 n_{\bar{\Delta}} - 2 n_\Delta - n_{{\tilde e}^c} .
\end{equation}
%%%%%

During the inflation the AD-flatons settle into one of the minima of $ V $,
%%%%%%
\begin{equation}
| \phi_a | \sim ( M / \lambda )^{1/2} H_{\rm inf}^{1/2} ,
\label{phi-inf}
\end{equation}
%%%%%%
where $ \lambda $ represents the mean value
of $ \lambda_{L \llap /} $ and $ \lambda_\Delta $.
The phases $ \theta_a $ of AD-flatons are fixed
with constant $ H_{\rm inf} $,
and the number asymmetries $ n_a $ are vanishing.
After the end of inflation, the AD-flatons evolve
with the decreasing Hubble parameter $ H = (2/3) t^{-1} $
in the matter-dominated universe as
%%%%%
\begin{equation}
| \phi_a | \sim {\bar \phi} \equiv ( M / \lambda )^{1/2} H^{1/2} .
\label{phi-0}
\end{equation}
%%%%%
Then, their phases $ \theta_a (t) $ begin to vary slowly in time
as $ | d \theta_a / d \ln t | \sim | \dot{\theta}_a | / H \lesssim 1 $,
since the balance among the phase-dependent terms in the scalar potential,
$ \lambda_{L \llap /} $-$ \lambda_\Delta $ cross term,
$ a_{L \llap /} $ term and $ a_\Delta $ term,
is changing with the decreasing Hubble parameter $ H $.
This causes the gradual fluctuation in $ \ln t $
of the fractions of number asymmetries,
$ \epsilon_a (t) \equiv n_a (t) / [ ( M / \lambda ) H^2 ] $,
which is not expected in the usual Affleck-Dine mechanism
along the one-dimensional flat direction.
Accordingly, even in this very early epoch
the lepton number asymmetry really appears
by this phase fluctuation of the AD-flatons on the flat manifold.
Numerically, we have
%%%%%
\begin{equation}
| \epsilon_L (t) | \lesssim 0.1 - 1 ,
\end{equation}
%%%%%
where
%%%%%
\begin{equation}
\epsilon_L (t) \equiv n_L (t) /  [ ( M / \lambda ) H^2 ]
\label{eL}
\end{equation}
%%%%%
represents the fraction of lepton number asymmetry.

%%%%%%%%%%%%%%%%%%%%
%%%%%%%%%%%%%%%%%%%%
\section{Completion of leptogenesis}
\label{sec:completion}
%%%%%%%%%%%%%%%%%%%%
%%%%%%%%%%%%%%%%%%%%

For the completion of leptogenesis,
the lepton number asymmetry should be fixed
to some significant value by the rapid redshift and oscillation
of the lepton-number violating terms.
In the original scenario \cite{PLB524},
it is found that the asymmetry is fixed readily in the early epoch
due to the effect of very large Higgs triplet mass terms
with $ M_\Delta \gtrsim 10^9 {\rm GeV} $,
which dominate fairly over the thermal terms
\cite{thermaleffect1,thermaleffect2}.
On the other hand, for the phenomenologically interesting case of
$ M_\Delta \sim 1 {\rm TeV} $,
the problem of asymmetry fixing becomes a more important issue
due to the effects of thermal terms.
We henceforth examine the case of $ M_\Delta \sim m_0 \sim  1 {\rm TeV} $
for the completion of leptogenesis.

The condensates of $ \Delta $ and $ {\tilde e}^c $
(more precisely $ \Delta^+ $ and $ {\tilde l}^c_1 $, respectively)
provide the effective superpotential mass terms
for $ L_i $ and $ H_d $ from the superpotential $ W_0 $
in Eq. (\ref{W0}) as
%%%%%
\begin{equation}
f_i l_i \nu_i \Delta
+ h_{i1} l_i H_d^0 {\tilde e}^c - h_{i1} H_d^- \nu_i {\tilde e}^c ,
\end{equation}
%%%%%
where $ l_i $, $ \nu_i $, $ H_d^0 $ and $ H_d^- $
are chiral superfields.
The lepton doublet basis is taken
with the diagonal $ f_{ij} = f_i \delta_{ij} $ ($ 0 < f_1 < f_2 < f_3 $).
The $ h $ couplings may be estimated
as $ | h_{i1} | \sim m_\tau / \langle H_d \rangle $
for the almost bi-maximal mixing of neutrinos.
We hence consider for definiteness the case
%%%%%
\begin{equation}
f_3 > f_2 \gtrsim 0.05 > f_1 \sim | h_{i1} | \sim 0.01 ,
\label{fhrange}
\end{equation}
%%%%%
which will be interesting phenomenologically
for the lepton flavor violation with the $ f $ couplings
\cite{t3ha,t3hb,t3hc,PRD69,rossi}.
In this case with the large $ | \phi_a | $ in Eq. (\ref{phi-0}),
the lepton doublets $ L_2 $ and $ L_3 $ acquire the masses
mainly with the $ f_2 $ and $ f_3 $ couplings, respectively,
while $ L_1 $ and $ H_d $ form a $ 2 \times 2 $ mass matrix
with the $ f_1 $ and $ h_{11} $ couplings in a good approximation.
Then, for a long period after the inflation
the lepton doublets $ L_i $, the Higgs doublet $ H_d $
and the gauge bosons $ W^\pm $ of $ {\rm SU(2)}_W / {\rm U(1)}_{I_3} $
and $ B $ of $ {\rm U(1)}_Y $ are heavy
enough to decouple from the dilute thermal plasma
of the inflaton decay products.
The plasma temperature before the reheating epoch of $ H = H_{\rm R} $
is given by
%%%%%
\begin{eqnarray}
T_{\rm p} \sim ( T_{\rm R}^2 H M_{\rm P} )^{1/4} \
(  H_{\rm R} < H < H_{\rm inf} ),
\label{Tp}
\end{eqnarray}
%%%%%
where $ M_{\rm P} = 2.4 \times 10^{18} $ GeV is the reduced Planck mass.
The reheating temperature is constrained
as $ T_{\rm R} \lesssim 10^7 {\rm GeV} $
to avoid the overproduction of gravitinos
with mass $ \sim m_0 \sim 1 {\rm TeV} $
\cite{gravitino,gravitino2}.

In this situation, the thermal-log terms appear
through the modification of the gauge coupling $ g_2 $
of unbroken $ {\rm U(1)}_{I_3} $ $ \subset {\rm SU(2)}_W $
due to the decoupling of $ W^\pm $, $ L_i $ and $ H_d $:
%%%%%
\begin{eqnarray}
V_{\rm thlog} &=& a_g \alpha_2^2 T_{\rm p}^4 {\rm ln}
\left[ ( | \Delta |^2 + | {\bar \Delta} |^2 ) / T_{\rm p}^2 \right]
\nonumber \\
&{}& + a_L \alpha_2^2 T_{\rm p}^4 {\rm ln}
\left[ | \Delta |^2 / T_{\rm p}^2 \right]
\nonumber \\
&{}& + a_H \alpha_2^2 T_{\rm p}^4 {\rm ln}
\left[ | {\tilde e}^c |^2 / T_{\rm p}^2 \right] ,
\label{Vthlog}
\end{eqnarray}
%%%%%
where
%%%%%
\begin{eqnarray}
a_g = - 6(27/64) , \ a_L = 27/64 , \ a_H = 27/64 ,
\end{eqnarray}
%%%%%
and $ \alpha_2 = g_2^2 / 4 \pi \approx 1/30 $.
The leading contribution to the gauge coupling dependent part
of the free energy is calculated as $ F = (27/64) g_2^2 T_{\rm p}^4 $
in the supersymmetric $ {\rm U(1)}_{I_3} $ gauge theory,
by using the formula in the literature \cite{free-energy}
for the chiral superfields of quarks, Higgs doublet ($ H_u $)
and Higgs triplets.
It should be noted here that the thermal-log terms $ V_{\rm thlog} $
act in total as {\it negative} one ($ a_g + a_L + a_H < 0 $),
providing a significant effect on the evolution of AD-flatons
for $ H < M_\Delta $.

On the other hand, with the decreasing $ | \phi_a | $
the lepton doublet $ L_1 $ and Higgs doublet $ H_d $
enter the thermal plasma in a later epoch
satisfying the condition
$  f_1 | \Delta |, h_{11} | {\tilde e}^c | < T_{\rm p} $.
Then, $ \Delta $ and $ {\tilde e}^c $ acquire
the thermal mass terms through the couplings with $ L_1 $ and $ H_d $ as
%%%%%
\begin{equation}
V_{\rm thm}
= \frac{1}{4} f_1^2 T_{\rm p}^2 | \Delta |^2
+ \frac{1}{2} h_{11}^2 T_{\rm p}^2 | {\tilde e}^c |^2 .
\label{VthD}
\end{equation}
%%%%%
The heavier lepton doublets $ L_2 $ and $ L_3 $
may also enter the thermal plasma with smaller $ | \Delta | $,
providing the similar mass terms.
These thermal mass terms may also make some effects
on the evolution of AD-flatons for $ H < M_\Delta $.

%%%%%%%%%%
\subsection{Case of $ M_\Delta > H_{\rm th} $}
%%%%%%%%%%

The AD-flatons are scaled as $ {\bar \phi} \propto H^{1/2} $
for some period after the inflation, as seen in Eq. (\ref{phi-0}).
Then, the Higgs triplet mass terms and the thermal-log terms,
scaled as $ H $, eventually become important
for the dynamics of AD-flatons.
As for the thermal mass terms $ V_{\rm thm} $
with the $ f $ and $ h $ couplings in Eq. (\ref{fhrange}),
they appear really after the thermal-log terms $ V_{\rm thlog} $
become dominant, as will be seen later.
We here consider specifically the case
that the Higgs triplet mass terms first dominate
over the Hubble induced mass terms
for $ H \sim M_\Delta \sim m_0 $ under the condition
%%%%%
\begin{eqnarray}
M_\Delta > H_{\rm th}
\sim {\sqrt a} \alpha_2 T_{\rm R} [ M_{\rm P} / ( M / \lambda ) ]^{1/2}
\label{MDdom}
\end{eqnarray}
%%%%%
with $ a = | a_g + a_L + a_H | = 27/16 $.
Here, the Hubble parameter $ H_{\rm th} $
is given by the condition $ H^2 | \phi_a |^2 \sim | V_{\rm thlog} | $
with Eqs. (\ref{phi-0}) and (\ref{Tp}),
for which the thermal-log terms would become
comparable to the Hubble induced mass terms
if the mass terms with $ M_\Delta \sim m_0 $ were subdominant.
It is estimated as
%%%%%
\begin{equation}
H_{\rm th} \sim 10^2 {\rm GeV}
\left( \frac{T_{\rm R}}{10^6 {\rm GeV}} \right)
\left( \frac{M / \lambda}{10^{23} {\rm GeV}} \right)^{-1/2} .
\label{Hth}
\end{equation}
%%%%%
Hence, the condition (\ref{MDdom})
for the dominance of the Higgs triplet mass terms
can be satisfied even for $ M_\Delta \sim 1 {\rm TeV} $
with relatively low reheating temperature,
which is favorable for avoiding the gravitino problem
\cite{gravitino,gravitino2}.

In this case of $ M_\Delta > H_{\rm th} $,
the AD-flatons begin to rotate for $ H \sim M_\Delta $
with frequency $ \sim M_\Delta \sim m_0 $ driven by the mass terms
\cite{PLB524}.
The AD-flatons are hence redshifted rapidly by rotation
for $ H \lesssim M_\Delta $ as
%%%%%
\begin{equation}
| \phi_a | \sim [ ( M / \lambda ) / M_\Delta ]^{1/2} H .
\label{phi-1}
\end{equation}
%%%%%
Then, after a while
the thermal-log terms $ \propto T_{\rm p}^4 \propto H $
catch up the mass terms $ \propto | \phi_a |^2 \propto H^2 $
for the Hubble parameter
%%%%%
\begin{equation}
H \sim H_{\rm th}^\prime
\sim ( H_{\rm th} / M_\Delta ) H_{\rm th}
= ( H_{\rm th} / M_\Delta )^2 M_\Delta .
\end{equation}
%%%%%
Accordingly, some minima are formed
by the effect of the negative thermal-log term as
%%%%%
\begin{eqnarray}
\phi_a (1) &:&
| \Delta | \sim | {\bar \Delta} | \sim | {\tilde e}^c |
\sim {\bar \phi}_{\rm th} ,
\label{minimum1}
\\
\phi_a (2) &:&
| \Delta | \simeq | {\bar \Delta} | \sim {\bar \phi}_{\rm th} ,
| {\tilde e}^c | = 0 ,
\label{minimum2}
\\
\phi_a (3) &:&
| \Delta | = 0 ,
| {\bar \Delta} | \simeq | {\tilde e}^c |
\sim {\bar \phi}_{\rm th}
\label{minimum3}
\end{eqnarray}
%%%%%
with
%%%%%
\begin{equation}
{\bar \phi}_{\rm th} = {\sqrt a} \alpha_2 T_{\rm p}^2 / M_\Delta
\label{phi-th}
\end{equation}
%%%%%
from the condition $ M_\Delta^2 | \phi_a |^2 \sim | V_{\rm thlog} | $.
In particular, before the reheating
%%%%%
\begin{equation}
{\bar \phi}_{\rm th} \sim ( H_{\rm th} / M_\Delta ) {\bar \phi}
\propto H^{1/2} \ ( H_{\rm R} < H \lesssim H_{\rm th}^\prime )
\label{phi-th1}
\end{equation}
%%%%%
with Eq. (\ref{Tp}) for $ T_{\rm p} $
and Eq. (\ref{MDdom}) for $ H_{\rm th} $.
The main terms to determine these minima are given by
%%%%%
\begin{eqnarray}
V_1
&=& V_{\rm thlog}
+ g_1^2 (|\Delta|^2 - |{\bar \Delta}|^2 + |{\tilde e}^c|^2)^2
\nonumber \\
&& + ( M_\Delta^2 + C_1 m_0^2 ) | \Delta |^2
+ ( M_\Delta^2 + C_2 m_0^2 ) | {\bar \Delta} |^2
\nonumber \\
&& + \left[ B_\Delta m_0 M_\Delta {\bar \Delta} \Delta
 + {\rm h.c.} \right] 
 + C_3 m_0^2 | {\tilde e}^c |^2 .
\label{V1}
\end{eqnarray}
%%%%%
It is really checked that
the thermal mass terms $ V_{\rm thm} $ do not appear in this epoch,
since $ f_1 | \Delta | \sim h_{11} | {\tilde e}^c | > T_{\rm p} $
for $ H \sim H_{\rm th}^\prime $
with $ f_1 \sim h_{11} \sim 0.01 $
and the reasonable range of $ T_{\rm R} \sim 10^5 - 10^7 {\rm GeV} $
and $ M / \lambda \lesssim 10^{23} {\rm GeV} $.
Here, it should be noticed that $ V_1 $ is degenerate
along the circles with radii $ \sim {\bar \phi}_{\rm th} $,
including the minima $ \phi_a (K) $ ($ K = 1 , 2 , 3 $),
in the complex planes of $ \phi_a $ under the constraint
$ \theta_\Delta + \theta_{\bar \Delta} + \arg ( B_\Delta ) = \pi $
for $ K = 1 , 2 $ to minimize the $ B_\Delta $ term
as $ - | B_\Delta | m_0 M_\Delta | {\bar \Delta} | | \Delta | $.
This degeneracy is slightly lifted by the higher order terms
in the whole potential, determining the phases of $ \phi_a (K) $
to form actually the minima.

Since the AD-flatons get significant angular momenta
by the effect of the mass terms with $ M_\Delta \sim m_0 $
under the condition (\ref{MDdom}),
they continue to rotate in the epoch of $ H < M_\Delta $.
Specifically, $ {\tilde e}^c $ may rotate almost freely
and redshifted as $ | {\tilde e}^c | \propto H $ toward the origin,
separated from $ \Delta $ and $ {\bar \Delta} $.
On the other hand, $ \Delta $ and $ {\bar \Delta} $
rotate around the minimum $ \phi_a (2) $
linked by the $ B_\Delta $ term and the $ D^2 $ term
with $ | \Delta | \simeq | {\bar \Delta} | \gg | {\tilde e}^c | $
after the thermal-log terms dominate over the Higgs triplet mass terms.
That is, for $ H \lesssim H_{\rm th}^\prime $
%%%%%
\begin{equation}
| \Delta | \simeq | {\bar \Delta} |
\sim {\bar \phi}_{\rm th} , \
| {\tilde e}^c | \sim [ ( M / \lambda ) / M_\Delta ]^{1/2} H .
\label{case-A}
\end{equation}
%%%%%
It seems rather unlikely that
the AD-flatons are trapped by the minimum $ \phi_a (1) $ or $ \phi_a (3) $,
once $ | {\tilde e}^c | \propto H $ is reduced suffciently
until $ H \sim H_{\rm th}^\prime $.

According to this redshift and rotation of AD-flatons,
the lepton number asymmetry is fixed
\cite{PLB524} as
%%%%%
\begin{equation}
\epsilon_L (t) \approx \epsilon_L \sim 0.1 \
( t \gg H_{\rm th}^{\prime -1} ) .
\label{eL1}
\end{equation}
%%%%%
The fixing of lepton number asymmetry can really be approved
by considering the rate equation,
%%%%%
\begin{eqnarray}
\frac{d \epsilon_L}{dt}
&=& - \frac{2}{H^2 ( M / \lambda )} \sum_a Q_L (a) {\rm Im} 
\left[ \phi_a \frac{\partial V}{\partial \phi_a} \right]
\nonumber \\
& \simeq &
- \frac{2 ( \lambda_{L \llap /} / \lambda )}{H^2 ( M / \lambda )^2}
{\rm Im} \left[ 2 M_\Delta \Delta^* {\bar \Delta} {\tilde e}^c {\tilde e}^c
\right]
\nonumber \\
&& - \frac{2 ( \lambda_{L \llap /} / \lambda )}{H^2 ( M / \lambda )^2}
{\rm Im} \left[ A_{L \llap /} m_0
{\bar \Delta} {\bar \Delta} {\tilde e}^c {\tilde e}^c \right] .
\label{deL/dt}
\end{eqnarray}
%%%%%
(The thermal terms $ V_{\rm thlog} + V_{\rm thm} $
conserve the particle numbers.)
The lepton-number violating sources in the right side of Eq. (\ref{deL/dt})
are given roughly as
$ ( H_{\rm th} / M_\Delta )^2 H \propto t^{-1}  $
with Eqs. (\ref{phi-th1}) and (\ref{case-A}),
and oscillate around zero with frequency $ \sim m_0 $
particularly due to the rapid rotation of $ {\tilde e}^c $.
Hence, the lepton number asymmetry is fixed to some significant value
for $ t \gg H_{\rm th}^{\prime -1} $
upon integration of Eq. (\ref{deL/dt}) in time.

%%%%%%%%%%%%%%%%%%%%
\begin{figure}[thb]
\begin{center}
\scalebox{.7}{\includegraphics*[3cm,2cm][15cm,28.5cm]{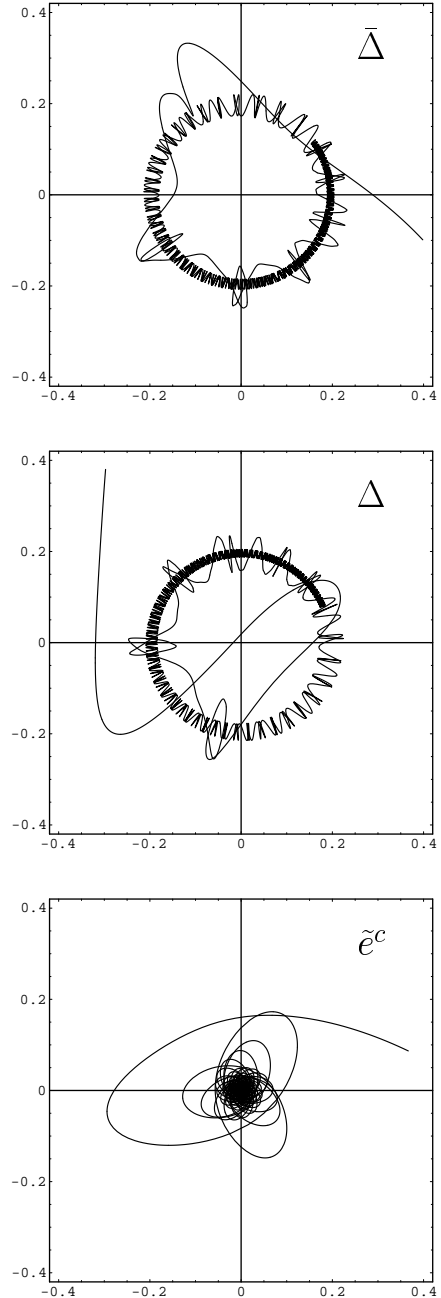}}
\caption{The evolution of AD-flatons  $ \phi_a / {\bar \phi} $ is shown
for $ M_\Delta^{-1} \lesssim t \lesssim 10 H_{\rm th}^{\prime -1} $
in the case of $ M_\Delta > H_{\rm th} $.
}
\label{tra}
\end{center}
\end{figure}
%%%%%%%%%%%%%%%%%%%%

%%%%%%%%%%%%%%%%%%%%
\begin{figure}[thb]
\begin{center}
\scalebox{.7}{\includegraphics*[0cm,14cm][14cm,22.5cm]{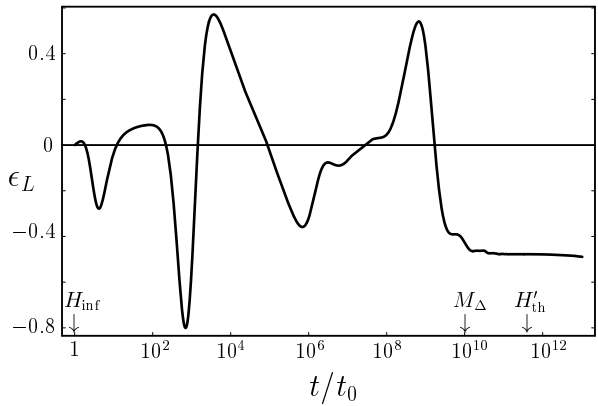}}
\caption{The time variation of the lepton number asymmetry is shown
for $ H_{\rm inf}^{-1} \lesssim t \lesssim 10 H_{\rm th}^{\prime -1} $
in the case of $ M_\Delta > H_{\rm th} $.
}
\label{e_l}
\end{center}
\end{figure}
%%%%%%%%%%%%%%%%%%%%

These arguments for the case of $ M_\Delta > H_{\rm th} $
on the evolution of AD-flatons and the lepton number asymmetry
($ t \sim H_{\rm inf}^{-1} \rightarrow t \gg H_{\rm th}^{\prime -1} $)
can be confirmed by numerical calculations.
A typical example is presented in Figs. \ref{tra} and \ref{e_l}
with $ M_\Delta = m_0 = 10^3 {\rm GeV} $,
$ T_{\rm R} = 5 \times 10^5 {\rm GeV} $,
$ M / \lambda = 2 \times 10^{22} {\rm GeV} $,
$ H_{\rm inf} = 10^{13} {\rm GeV} $
($ H_{\rm th} \sim 10^2 {\rm GeV} $,
$ H_{\rm th}^\prime \sim 10 {\rm GeV} $)
and certain values of the other parameters in the reasonable range.
It is clearly seen in Fig. \ref{tra}
that the trajectories of $ \Delta $ and $ {\bar \Delta} $
in unit of $ {\bar \phi} \propto H^{1/2} $
with Eq. (\ref{phi-th1}) are converging to the circle with radius
$ \sim H_{\rm th} / M_\Delta \sim 0.1 $ satisfying
$ | \Delta | \simeq | {\bar \Delta} | \sim {\bar \phi}_{\rm th} $.
The trajectory of $ {\tilde e}^c $ is, on the other hand,
shrinking as $ | {\tilde e}^c | / {\bar \phi} \propto H^{1/2} $.
It is also seen in Fig. \ref{e_l}
that the lepton number asymmetry $ \epsilon_L (t) $ is fixed
for $ t \gg H_{\rm th}^{\prime -1} $.

In this way, the lepton-to-entropy ratio at the reheating is estimated
from Eqs. (\ref{eL}), (\ref{eL1})
and the entropy density $ s \approx 4 H^2 M_{\rm P}^2 / T_{\rm p} $
with $ H = H_{\rm R} $ and $ T_{\rm p} = T_{\rm R} $ as
%%%%%
\begin{equation}
\frac{n_L}{s}
\sim 10^{-10}
    \left( \frac{\epsilon_L}{0.1} \right)
    \left( \frac{T_{\rm R}}{10^6 {\rm GeV}} \right) 
    \left( \frac{M / \lambda}{10^{22} {\rm GeV}} \right) .
\label{nL/s1}
\end{equation}
%%%%%
The Hubble parameter in the radiation-dominated epoch is given by
%%%%%
\begin{equation}
H = ( \pi / {\sqrt{90}} ) {\sqrt{g_*}} T_{\rm p}^2 / M_{\rm P} \
( H \lesssim H_{\rm R} )
\label{Hrd}
\end{equation}
%%%%%
with $ g_* \approx 200 $.
It should be noted here that for a long period of $ H < H_{\rm th}^\prime $
the lepton number asymmetry is still stored
in the condensates of AD-flatons rotating around the potential trap
$ \phi_a (2) $ formed by the negative thermal-log term.
While this situation may continue even after the reheating,
the lepton-to-entropy ratio $ n_L / s $
as given in Eq. (\ref{nL/s1}) remains constant
(without significant extra entropy production).
This is approved by considering the rate equation for $ n_L / s $
with $ s \propto H^{3/2} $ in the radiation-dominated universe,
which is similar to Eq. (\ref{deL/dt}).

The AD-flatons should be liberated anyway from the potential trap
to complete the leptogenesis
while the sphaleron process is effective
to convert the lepton number to the baryon number.
This liberation takes place when the negative thermal-log term disappears
by the thermalization of the gauge bosons $ W^\pm $.
After the thermal-log terms $ V_{\rm thlog} $
become dominant for $ H \sim H_{\rm th}^\prime $,
the Higgs doublet $ H_d $ is first thermalized
with $ h_{11} \sim f_1 \sim 0.01 $
since $ | {\tilde e}^c | \propto H $ is reduced faster
than $ | \Delta | \propto H^{1/2} $, as seen in Eq. (\ref{case-A}).
Then, after a while the lepton doublet $ L_1 $
also enters the thermal plasma,
providing the thermal mass terms $ V_{\rm thm} $.
This may occur before or after the reheating
depending on whether $ M_\Delta > f_1 {\sqrt a} \alpha_2 T_{\rm R} $
or $ M_\Delta < f_1 {\sqrt a} \alpha_2 T_{\rm R} $
with $ f_1 {\sqrt a} \alpha_2 \sim 10^{-3} $, as seen below.
The relevant Hubble parameter is estimated
from the condition $ f_1 | \Delta | \sim T_{\rm p} $
by considering $ | \Delta | \simeq | {\bar \Delta} |
\sim {\bar \phi}_{\rm th} $ with Eq. (\ref{phi-th})
for $ {\bar \phi}_{\rm th} $
and Eqs. (\ref{Tp}) and (\ref{Hrd}) for $ T_{\rm p} $
as
%%%%%
\begin{equation}
H_{\rm thm} \sim \left\{ \begin{array}{ll}
H_{\rm R} ( M_\Delta / f_1 {\sqrt a} \alpha_2 T_{\rm R} )^4
& ( H_{\rm thm} > H_{\rm R} ) \\
H_{\rm R} ( M_\Delta / f_1 {\sqrt a} \alpha_2 T_{\rm R} )^2
& ( H_{\rm thm} < H_{\rm R} ) \end{array} \right. .
\end{equation}
%%%%%
Here, the Hubble parameter $ H_{\rm R} $ at the reheating
is estimated with Eq. (\ref{Hrd}) as
%%%%%
\begin{equation}
H_{\rm R} \sim 10^{-6} {\rm GeV}
\left( \frac{T_{\rm R}}{10^6 {\rm GeV}} \right)^2 .
\label{HR}
\end{equation}
%%%%%
By considering the condition
$ f_1 | \Delta | \sim f_1 {\bar \phi}_{\rm th} \sim T_{\rm p} $
with Eq. (\ref{phi-th}),
we have a relation $ f_1 T_{\rm p} \sim T_{\rm p}^2 / {\bar \phi}_{\rm th}
\sim M_\Delta / {\sqrt a} \alpha_2 $ for $ H \sim H_{\rm thm} $.
That is, the thermal mass $ (1/2) f_1 T_{\rm p} $
is fairly larger than $ M_\Delta $
by a factor $ \sim 1 / ( 2 {\sqrt a} \alpha_2 ) \approx 10 $.
Hence, the potential minimum $ \phi_a (2) $ is shifted
by the balance between the thermal mass terms and the thermal-log terms as
%%%%%
\begin{eqnarray}
| \Delta | \simeq | \bar \Delta |
\simeq ( 2 {\sqrt a} \alpha_2 / f_1 ) T_{\rm p}
\sim ( 2 {\sqrt a} \alpha_2 ) {\bar \phi}_{\rm th}
\label{Delta-reduction}
\end{eqnarray}
%%%%%
with the reduction of $ | \Delta | \simeq | \bar \Delta | $
from $ {\bar \phi}_{\rm th} $
by a factor $ \sim 2 {\sqrt a} \alpha_2 \approx 0.1 $,
where the condition $ f_1 {\bar \phi}_{\rm th} \sim T_{\rm p} $
for $ H \sim H_{\rm thm} $ is considered.

It may be expected here that the $ f $ couplings satisfy the condition
%%%%%
\begin{equation}
f_i < f_{i+1} < f_i / ( 2 {\sqrt a} \alpha_2 ) \approx 10 f_i \
( i = 1 , 2 ) ,
\label{fcond1}
\end{equation}
%%%%%
which is consistent with the hierarchical neutrino mass spectrum,
e.g., $ f_1 = 0.01 $, $ f_2 = 0.05 $, $ f_3 = 0.3 $.
Then, since $ f_2 | \Delta |
\simeq 2 {\sqrt a} \alpha_2 ( f_2 / f_1 ) T_{\rm p} < T_{\rm p} $
from Eqs. (\ref{Delta-reduction}) and (\ref{fcond1}),
the second lepton doublet $ L_2 $ also enters the thermal plasma,
and by the effect of thermal mass term
$ (1/4) f_2^2 T_{\rm p}^2 | \Delta |^2 $
the Higgs triplets are reduced further
as $ ( 2 {\sqrt a} \alpha_2 / f_1 ) T_{\rm p}
\rightarrow ( 2 {\sqrt a} \alpha_2 / f_2 ) T_{\rm p} $.
In this way, when the lightest $ L_1 $ is thermalized
for $ H \sim H_{\rm thm} $,
the heavier $ L_2 $ and $ L_3 $ sequentially come into the thermal plasma,
providing larger thermal mass terms for $ \Delta $.
As a result, if the condition
%%%%%
\begin{equation}
f_i > g_2 ( 2 {\sqrt a} \alpha_2 ) \approx 0.07 \
({\mbox{$ i $ = 2 or 3}})
\end{equation}
%%%%%
is further satisfied,
the $ {\rm SU(2)}_W / {\rm U(1)}_{I_3} $ gauge bosons $ W^\pm $
with mass $ M_W
= ( g_2 / {\sqrt 2} ) ( | \Delta |^2 + | {\bar \Delta} |^2 )^{1/2}
\simeq g_2 ( 2 {\sqrt a} \alpha_2 / f_i ) T_{\rm p} < T_{\rm p} $
($ i $ = 2 or 3) may even be thermalized
soon after the lepton doublet $ L_1 $ enters the thermal plasma
for $ H \sim H_{\rm thm} $.

It is, on the other hand, considered that the above conditions
on $ f_i $ and $ g_2 $ may not be satisfied.
Then, since the thermal mass terms decrease with $ T_{\rm p} $,
the mass terms with $ M_\Delta \sim m_0 $ dominate again
after a while for $ H < H_{\rm thm} $
so that the minimum returns to $ \phi_a (2) $
from Eq. (\ref{Delta-reduction}).
Even in this case, the negative thermal-log term disappears
in a later epoch when the gauge bosons are thermalized
satisfying the condition $ M_W \simeq g_2 | \Delta | < T_{\rm p} $
($ | \Delta | \simeq | {\bar \Delta} | $).
The relevant Hubble parameter is estimated
with $ | {\bar \Delta} | \sim {\bar \phi}_{\rm th} \propto T_{\rm p}^2 $
in Eq. (\ref{phi-th}), $ T_{\rm p} $ in Eq. (\ref{Hrd})
and $ g_2^2 a \alpha_2^2 \approx 10^{-3} $ as
%%%%%
\begin{eqnarray}
H_{\rm thg}
& \sim & 10^{-9} {\rm GeV}
\left( \frac{M_\Delta}{10^3 {\rm GeV}} \right)^2 ,
\label{Hthg}
\end{eqnarray}
%%%%%
which is really smaller than $ H_{\rm R} $ in Eq. (\ref{HR}).

Once the negative thermal-log term disappears, as seen so far,
the minimum is moved to the origin $ \phi_a = 0 $.
Then, the condensates of AD-flatons
with energy densities $ \sim a \alpha_2^2 T_{\rm p}^4 $
($ \ll T_{\rm p}^4 $)
for $ \Delta $ and $ {\bar \Delta} $
and a smaller amount for $ {\tilde e}^c $
are evaporated through the lepton-number conserving gauge interactions
without significant entropy production.
Accordingly, the lepton number asymmetry stored
in the AD-flatons is released to the thermal plasma
through this evaporation process.
This occurs for $ H \sim H_{\rm thm} $ or $ H_{\rm thg} $,
which is fairly before the sphaleron process is freezed out
for the Hubble parameter $ H_{\rm sph} \sim 10^{-14} {\rm GeV} $
with $ T_{\rm p} \sim 10^2 {\rm GeV} $ of the electroweak phase transition.
Then, the lepton number asymmetry is finally converted
to the baryon number asymmetry
through the sphaleron process as $ n_B = - ( 8 / 23 ) n_L $
\cite{FY,hatu}.
Therefore, the sufficient baryon number asymmetry can be provided
for the nucleosynthesis with $ \eta = ( 6.1 \pm 0.2 ) \times 10^{-10} $
\cite{eta}
from the lepton number asymmetry as given in Eq. (\ref{nL/s1}).

%%%%%%%%%%
\subsection{Case of $ M_\Delta < H_{\rm th} $}
%%%%%%%%%%

We in turn examine the case of $ M_\Delta < H_{\rm th} $
with larger $ T_{\rm R} \sim 10^7 {\rm GeV} $
and smaller $ M / \lambda \sim 10^{20} {\rm GeV} $ in Eq. (\ref{Hth}),
where the evolution of AD-flatons appears substantially
different from that for the case of $ M_\Delta > H_{\rm th} $.
The negative thermal-log term first dominates
over the Hubble induced mass terms for $ H \sim H_{\rm th} $,
and it soon competes with the other terms in $ V_1 $,
forming the potential minima
in Eqs. (\ref{minimum1}), (\ref{minimum2}), (\ref{minimum3}).
The AD-flatons are already tracking the instantaneous minimum
$ | \phi_a | \sim {\bar \phi} $ with fluctuating phases after the inflation
\cite{PLB524}.
In this course, $ {\tilde e}^c $ is linked
with $ \Delta $ and $ {\bar \Delta} $ by the quartic terms
$ \Delta^* {\bar \Delta} {\tilde e}^c {\tilde e}^c $
and $ {\bar \Delta} {\bar \Delta} {\tilde e}^c {\tilde e}^c $
with couplings $ \sim M_\Delta / ( M / \lambda ) $ in $ V $.
Then, for $ H \lesssim H_{\rm th} $
the AD-flatons $ \Delta $, $ {\bar \Delta} $, $ {\tilde e}^c $
together move gradually toward the minimum $ \phi_a (1) $
due to the effect of the negative thermal-log term.

If the Hubble parameter decreases further to
$ H \sim ( M_{\Delta} / H_{\rm th} )^3 H_{\rm th} $,
the mass term $ m_0^2 | {\tilde e}^c |^2 \propto H $
dominates over the quartic terms
$ \Delta^* {\bar \Delta} {\tilde e}^c {\tilde e}^c $
and $ {\bar \Delta} {\bar \Delta} {\tilde e}^c {\tilde e}^c $
$ \propto H^2 $
to link $ {\tilde e}^c $ with $ \Delta $ and $ {\bar \Delta} $.
Then, $ {\tilde e}^c $ begins to oscillate
toward the origin with redshift faster as $ H $
rather than $ {\bar \phi}_{\rm th} \propto H^{1/2} $,
so that the AD-flatons turn to move
from the minimum $ \phi_a (1) $ to the minimum $ \phi_a (2) $
with $ | {\tilde e}^c | = 0 $.
Since the negative thermal-log term dominates
much earlier than the mass terms with $ M_\Delta \sim m_0 $,
the AD-flatons do not get significant angular momenta
until $ H \sim ( M_{\Delta} / H_{\rm th} )^3 H_{\rm th} $.
Hence, for $ H \lesssim ( M_{\Delta} / H_{\rm th} )^3 H_{\rm th} $
$ \Delta $ and $ {\bar \Delta} $ move gradually
toward the minimum $ \phi_a (2) $
without rotating around the circle including $ \phi_a (2) $.
On the other hand, $ {\tilde e}^c $ with small angular momentum
shows a complicated motion around the origin,
changing frequently the sign of the time derivative of its phase
$ {\dot \theta}_{{\tilde e}^c} $,
as seen by numerical calculations.
This would imply that $ {\tilde e}^c $ is not liberated fully
from $ \Delta $ and $ {\bar \Delta} $ in the presence of the $ D^2 $ term
and the phase-dependent quartic terms
$ \Delta^* {\bar \Delta} {\tilde e}^c {\tilde e}^c $
and $ {\bar \Delta} {\bar \Delta} {\tilde e}^c {\tilde e}^c $.

Particularly due to this complicated behavior of $ {\tilde e}^c $
linked to $ \Delta $ and $ {\bar \Delta} $
for the case of $ M_\Delta < H_{\rm th} $,
the lepton number asymmetry $ \epsilon_L (t) $
oscillates violently for $ H < H_{\rm th} $.
Its mean maginitude, on the other hand,
tends to be saturated to some large value
$ \epsilon_L \sim 10 - 10^2 $
for $ H < ( M_{\Delta} / H_{\rm th} )^3 H_{\rm th} $
after the epoch of transition from $ \phi_a (1) $ to $ \phi_a (2) $,
as seen by numerical calculations.
After the reheating, the plasma temperature $ T_{\rm p} $ decreases faster
as $ H^{1/2} $ rather than $ H^{1/4} $,
so that $ \Delta $ and $ {\bar \Delta} $ trapped
by the minimum $ \phi_a (2) $ also decrease faster
as $ T_{\rm p}^2 \propto H $.
Then, by considering the rate equation
it would be expected that the lepton number asymmetry is fixed,
providing a significant amount of $ n_L / s \sim 10^{-10} $
in Eq. (\ref{nL/s1}) with $ \epsilon_L \sim 10 $
for $ T_{\rm R} \sim 10^7 {\rm GeV} $
and $ M / \lambda \sim 10^{20} {\rm GeV} $.
In the case of $ M_\Delta < H_{\rm th} $, however,
it seems difficult to make a reliable estimate of the asymmetry
due to the substantial effect of the negative thermal-log term.

%%%%%%%%%%%%%%%%%%%%
%%%%%%%%%%%%%%%%%%%%
\section{Summary}
\label{sec:summary}
%%%%%%%%%%%%%%%%%%%%
%%%%%%%%%%%%%%%%%%%%

In summary, we have investigated the leptogenesis
with the supersymmetric Higgs triplets
in the light of experimental verification in the TeV region.
The lepton number asymmetry really appears after the inflation
via multiscalar coherent evolution of $ \Delta $, $ {\bar \Delta} $
and $ {\tilde e}^c $ on the flat manifold.
If the Higgs triplet mass terms dominate over the negative thermal-log term
for $ H \sim M_\Delta $, the asymmetry is fixed readily
to some significant value by the redshift and rotation of the AD-flatons,
providing the sufficient lepton-to-entroy ratio $ n_L / s \sim 10^{-10} $.
This can be the case even with $ M_\Delta \sim 1 {\rm TeV} $
for $ T_{\rm R} \sim 10^6 {\rm GeV} $
and $ M / \lambda \sim 10^{22} {\rm GeV} $.
On the other hand, if the negative thermal-log term dominates first,
the evolution of the AD-flatons appears different substantially.
Even in this case, a significant amount of $ n_L / s \sim 10^{-10} $
might be obtained for $ T_{\rm R} \sim 10^7 {\rm GeV} $
and $ M / \lambda \sim 10^{20} {\rm GeV} $,
though it seems difficult to make a reliable estimate of the asymmetry.

\acknowledgments

This work is supported in part by
Grant-in-Aid for Scientific Research on Priority Areas B (No. 13135214)
from the Ministry of Education, Culture, Sports, Science and Technology,
Japan.


\begin{thebibliography}{99}

\bibitem{higgstriplet}
J. Schechter and J. W. F. Valle,
Phys. Rev. D {\bf 22}, 2227 (1980);
R. N. Mohapatra and G. Senjanovi$\rm \acute{c}$,
Phys. Rev. Lett. {\bf 44}, 912 (1980);
Phys. Rev. D {\bf 23}, 165 (1981);
E. Ma and U. Sarkar,
Phys. Rev. Lett. {\bf 80}, 5716 (1998);

\bibitem{t3ha}
E. Ma and M. Raidal,
Phys. Rev. Lett. {\bf 87}, 011802 (2001);
{\it Erratum-ibid.} {\bf 87}, 159901 (2001).

\bibitem{t3hb}
E. J. Chun, K. Y. Lee, and S. C. Park,
Phys. Lett. B {\bf 565}, 142 (2003);
M. Kakizaki, Y. Ogura, and F. Shima,
Phys. Lett. B {\bf 565}, 210 (2003).

\bibitem{t3hc}
D. Aristizabal, M. Hirsch, J. W. F. Valle, and A. Villanova,
Phys. Rev. D {\bf 68}, 033006 (2003).

\bibitem{PRD69}
M. Senami and K. Yamamoto,
Phys. Rev. D {\bf 69}, 035004 (2004).

\bibitem{rossi}
A. Rossi,
Phys. Rev. D {\bf 66}, 075003 (2002).

\bibitem{PLB524}
M. Senami and K. Yamamoto,
Phys. Lett. B {\bf 524}, 332 (2002).

\bibitem{PRD66-67}
M. Senami and K. Yamamoto,
Phys. Rev. D {\bf 66}, 035006 (2002);
{\it ibid.} {\bf 67}, 095005 (2003).

\bibitem{AD}
I. Affleck and M. Dine,
Nucl. Phys. {\bf B 249}, 361 (1985).

\bibitem{DRT}
M. Dine, L. Randall and S. Thomas,
Phys. Rev. Lett. {\bf 75}, 398 (1995).
M. Dine, L. Randall and S. Thomas,
Nucl. Phys. {\bf B 458}, 291 (1996).

\bibitem{thermaleffect1}
R. Allahverdi, B. A. Campbell, and J. Ellis,
Nucl. Phys. B {\bf 579}, 355 (2000).

\bibitem{thermaleffect2}
A. Anisimov and M. Dine,
Nucl. Phys. B {\bf 619}, 729 (2001);
M. Fujii, K. Hamaguchi, and T. Yanagida,
Phys. Rev. D {\bf 63}, 123513 (2001).

\bibitem{SK}
Super-Kamiokande Collaboration, S. Fukuda et al.,
Phys. Rev. Lett. {\bf 86}, 5651 (2001);
{\bf 86}, 5656 (2001);
Phys. Lett. B {\bf 539}, 179 (2002).

\bibitem{SNO}
SNO Collaboration, Q. R. Ahmad et al.,
Phys. Rev. Lett. {\bf 87} (2001) 071301;
{\bf 89}, 011302 (2002); {\bf 92}, 181301 (2004).

\bibitem{CHOOZ}
CHOOZ Collaboration, M. Apollonio et al.,
Phys. Lett. B {\bf 466}, 415 (1999).

\bibitem{KamLand}
KamLAND Collaboration, K. Eguchi et al.,
Phys. Rev. Lett. {\bf 90}, 021802 (2003).

\bibitem{extra-dimension}
E. Ma, M. Raidal, and U. Sarkar, Phys. Rev. Lett. {\bf 85}, 3769 (2000);
Nucl. Phys. {\bf B 615}, 313 (2001).

\bibitem{natural}
't Hooft, {\it Recent Developments in Gauge Theories},
Proc. of Nato Advanced Study Institute (Carg{\`e}se, 1979),
Plenum, New York, 1980.

\bibitem{flaton}
K. Yamamoto,
Phys. Lett. B {\bf 161}, 289 (1985);
{\it ibid.} B {\bf 168}, 341 (1986).

\bibitem{free-energy}
J. Grundberg, T. H. Hansson and U. Lindstr\"om,
hep-th/9510045 (1995).

\bibitem{gravitino}
J. Ellis, J. E. Kim, and D. V. Nanopolous,
Phys. Lett. {\bf B 145}, 181 (1984);
E. Holtmann, M. Kawasaki, K. Kohri, and T. Moroi,
Phys. Rev. D {\bf 60}, 023506 (1999);
M. Kawasaki, K. Kohri, and T. Moroi,
{\it ibid.} {\bf 63}, 103502 (2001);
K. Kohri,
{\it ibid.} {\bf 64}, 043515 (2001);
R. H. Cyburt, J. Ellis, B. D. Fields, and K. A. Olive,
{\it ibid.} {\bf 67}, 103521 (2003);
M. Kawasaki, K. Kohri, and T. Moroi, Phys. Lett. B {\bf 625}, 7 (2005).

\bibitem{gravitino2}
G.F. Giudice, I. Tkachev, and A. Riotto,
JHEP {\bf 08}, 009 (1999); {\bf 11}, 036 (1999);
R. Kallosh, L. Kofman, A. Linde, and A. Van Proeyen,
Phys. Rev. D {\bf 61}, 103503 (2000);
A. L. Maroto and A. Mazumdar,
Phys. Rev. Lett. {\bf 84}, 1655 (2000);
R. Allahverdi, M. Bastero-Gil, and A. Mazumdar,
Phys. Rev. D {\bf 64}, 023516 (2001);
H. P. Nilles, M. Peloso, and L. Sorbo,
Phys. Rev. Lett {\bf 87}, 051302 (2001).

\bibitem{FY}
M. Fukugita and T. Yanagida,
Phys. Lett. B {\bf 174}, 45 (1986).

\bibitem{hatu}
J. A. Harvey and M. S. Turner,
Phys. Rev. D {\bf 42}, 3344 (1990).

\bibitem{eta}
S. Eidelman et al., Particle Data Group,
Phys. Lett. B {\bf 592}, 1 (2004), http://pdg.lbl.gov/.


\end{thebibliography}
\end{document}